\documentclass[fleqn, 11pt, nofootinbib, preprint]{revtex4}
\usepackage[dvipdfm]{graphicx}
\usepackage{amsmath,amssymb}
\usepackage{bm}
\usepackage{color}
\begin{document}

\title{Lepton flavor violation in the supersymmetric seesaw model \\ after the LHC 8 TeV run}
\date{December 8, 2014}

\author{Toru GOTO}
\email{tgoto@post.kek.jp}
\preprint{KEK-TH-1774}

\author{Yasuhiro OKADA}
\email{yasuhiro.okada@kek.jp}
\affiliation{Theory center, IPNS, KEK, 1-1 Oho, Tsukuba, Ibaraki, 305-0801, Japan \\
Department of Particle and Nuclear Physics, Graduate University for Advanced Studies (Sokendai), 1-1 Oho, Tsukuba, Ibaraki, 305-0801, Japan}

\author{Tetsuo SHINDOU}
\email{shindou@cc.kogakuin.ac.jp}
\affiliation{Division of Liberal Arts, Kogakuin University, 1-24-2 Shinjuku, Tokyo 163-8677, Japan}
\preprint{KU-PH-016}

\author{Minoru TANAKA}
\email{tanaka@phys.sci.osaka-u.ac.jp}
\affiliation{Department of Physics, Graduate School of Science, Osaka University, Toyonaka, Osaka 560-0043, Japan}
\preprint{OU-HET 828}

\author{Ryoutaro WATANABE}
\email{wryou1985@ibs.re.kr}
\affiliation{Center for Theoretical Physics of the Universe, Institute for Basic Science (IBS), Daejeon, 305-811, Republic of Korea}
\preprint{CTPU-14-09}

\begin{abstract}
We study the lepton flavor violation in the supersymmetric seesaw model 
taking into account recent experimental improvements, especially for the Higgs boson mass measurement, direct searches of superpartners and the rare decay of $B_s \to \mu^+\mu^-$ at the LHC, 
the neutrino mixing angle of $\theta_{13}$ at the neutrino experiments, and the search of $\mu\to e\gamma$ at the MEG experiment.   
We obtain the latest constraints on the parameters in the supersymmetry breaking terms  
and study the effect on the lepton flavor violating decays of $\tau\to\mu\gamma$ and $\mu\to e\gamma$. 
In particular, we consider two kinds of assumption on the structures in the Majorana mass matrix and the neutrino Yukawa matrix. 
In the case of the Majorana mass matrix proportional to the unit matrix, allowing non-vanishing CP violating parameters in the neutrino Yukawa matrix, 
we find that the branching ratio of $\tau\to\mu\gamma$ can be larger than $10^{-9}$ within the improved experimental limit of $\mu\to e\gamma$. 
We also consider the neutrino Yukawa matrix that includes the mixing only in the second and third generations, 
and find that a larger branching ratio of $\tau\to\mu\gamma$ than $10^{-9}$ is possible with satisfying the recent constraints. 
\end{abstract}

\maketitle

\section{Introduction}
The discovery of neutrino oscillations~\cite{NuOscillation} means that lepton flavors are not conserved and that the minimal standard model (SM) with massless neutrinos must be extended.
The seesaw mechanism is a simple and attractive extension to introduce the neutrino masses \cite{SeeSaw}. 
Similarly to the quark sector in the SM, a new additional Yukawa matrix for right-handed neutrinos induces lepton flavor violations (LFVs) in the charged lepton sector.  
In this simple extension, however, the LFV processes only occur via loop including a neutrino and are suppressed by small neutrino masses.  
If this is the case, it is practically impossible to observe the LFV except the neutrino oscillations.

An interesting extension is to impose supersymmetry (SUSY) on the seesaw mechanism~\cite{SUSYseesaw}. 
The LFV processes are induced via loop contributions from charged sleptons and sneutrinos whose masses are expected not to be far away from the electroweak scale. 
There is a possibility to enhance the LFV processes so that they can be measured at near future experiments.

In recent years, there are great experimental developments and then the allowed ranges of the model parameters has significantly changed.  
The most important development is that a Higgs boson was discovered from a diphoton decay \cite{ATLAShiggsObs, CMShiggsObs} 
in July 2012 at the Large Hadron Collider (LHC), whose mass is now found to be about $126\,\text{GeV}$ \cite{ATLAShiggs, CMShiggs}. 
Superpartners of the SM particles have not been discovered yet and the lower mass bound of colored superpartners is about 1~TeV after 8~TeV run of the LHC \cite{ATLASsusy, CMSsusy}.  
Flavor experiments also give us improved constraints on new physics.  
For $B$ physics, the recent improvement of $\mathcal B(b\to s\gamma)$ and the evidence of $B_s \to \mu^+\mu^-$ have impacts on the model~\cite{bsgHFAG, BsMuMuLHCbFirst, BsMuMuComb, BsMuMuLHCb, BsMuMuCMS}. 
For the lepton sector, the neutrino mixing angles introduced in the Pontecorvo-Maki-Nakagawa-Sakata (PMNS) matrix~\cite{PMNS} have been precisely determined 
by many kinds of neutrino experiments summarized in Ref.~\cite{PDGneutrino}, 
including the recent improvements~\cite{Theta13T2K,Theta13MINOS,Theta13Chooz,Theta13Daya,Theta13RENO} for $\sin \theta_{13}$. 
In addition, bounds on the branching ratios of LFV processes for $\mu\to e\gamma$ \cite{MuEgMEG} and $\tau \to (\mu,e)\gamma$ \cite{TauMuEbabar,TauMuEbelle} have been strengthened.

As a consequence of the Higgs boson discovery and the limit on SUSY particle masses, we expect that a scale of SUSY breaking is very high, 
or $A$-term, a trilinear scalar interaction term in the stop sector, is tuned to reproduce the correct Higgs boson mass.
In the supersymmetric seesaw model, LFV processes depend on both the structure of the neutrino sector and SUSY model parameters. 
In the previous work \cite{GOST} in 2008, the supersymmetric seesaw model of type I was studied. 
In general $\tau\to\mu\gamma$ is severely constrained by the experimental bound on $\mu\to e\gamma$ because their decay branching ratios are related,    
but it is shown that the ratio $\mathcal B(\tau\to\mu\gamma)/\mathcal B (\mu\to e\gamma)$ can be enhanced in several cases of the flavor structure of the seesaw sector. 
For example, a large enhancement of the ratio could occur 
with the simplest structure that the Majorana mass matrix is proportional to the unit matrix and the neutrino Yukawa matrix is real 
provided that the neutrino mixing angle of $\theta_{13}$ is zero and the neutrino masses are inversely hierarchical. 
In addition, such an enhancement is also found if the neutrino Yukawa matrix is assumed to have a mixing only between the second and third generations for both cases of the normal and inverted hierarchies of the neutrino masses. 
Another enhancement mechanism is also discussed in Ref.~\cite{ShinPetc} by considering the effect of the CP violating parameters in the neutrino Yukawa matrix. 
There are many other studies for the LFV in the SUSY seesaw models~\cite{PreviousStudy}, 
taking several constraints of the day into account. 
However, these studies have to be re-examined due to the above experimental improvements. 
Recent studies have been done for $\mu\to e\gamma$ in the case of a simple assumption on the neutrino Yukawa matrix and Majorana mass matrix 
in Refs.~\cite{RecentStudy,Figueiredo:2013tea}. 
The tau and muon LFV decays have recently been studied in the model embedded in an SO(10) grand unified theory (GUT)~\cite{Calibbi:2012gr}.

In the present work, we investigate both tau and muon LFV decays in the supersymmetric seesaw model of type I. 
We take the universal soft SUSY breaking and assume several specific structures on the neutrino Yukawa matrix and the Majorana mass matrix. 
In order to find how much the latest constraints change the previous results in Refs.~\cite{GOST,ShinPetc}, we first determine allowed regions of SUSY parameters. 
Then, we analyze the LFV decays $\tau\to\mu\gamma$ and $\mu\to e\gamma$.  
As a result, we see that the enhancement of $\mathcal B(\tau\to\mu\gamma)/\mathcal B (\mu\to e\gamma)$ is unlikely to occur 
for the simplest structure with the degenerate Majorana mass matrix and the real neutrino Yukawa matrix 
because of the sizable value of $\theta_{13}$. 
On the other hand, in the case of degenerate Majorana mass together with CP violating parameters in the neutrino Yukawa matrix, 
we find that an enhancement of $\mathcal B(\tau\to\mu\gamma)/\mathcal B (\mu\to e\gamma)$ is possible.  
We also consider the neutrino Yukawa matrix that includes the mixing only in the second and third generations, 
and find that the branching ratio of $\tau\to\mu\gamma$ can be as large as $10^{-9}$.  
These results imply that there is a good possibility for the tau LFV decay to be measured at the SuperKEKB/Belle~II~\cite{Aushev:2010bq} experiment 
in addition to the muon LFV decay at the upgraded MEG experiment (MEG~II)~\cite{Baldini:2013ke} .

This article is organized as follows. 
In Sec.~\ref{Ssm}, we review the supersymmetric seesaw model and summarize parametrizations for the seesaw sector. 
We describe our analysis method to evaluate flavor signals in Sec.~\ref{Am}. 
We present numerical results in Sec.~\ref{Nr}. 
Summary and conclusion are given in Sec.~\ref{Co}.

\section{Supersymmetric seesaw model}
\label{Ssm}

\subsection{Overview of the model}\label{SubSec:seesaw model}
In this section, we briefly review the supersymmetric seesaw model and summarize its features.
As is well known, the seesaw mechanism describes the tiny neutrino masses by introducing a new high mass scale. 
In the case of the type-I seesaw model, such a high scale is identical to the right-handed neutrino mass scale.
A minimal supersymmetric version of the type-I seesaw model is defined by a superpotential as 
\begin{eqnarray}
 \label{Eq:SuperLepton}
 W_\text{lepton} = 
 Y_E^{ij} E^c_i L_j H_1 +  Y_N^{ij} N^c_i L_j H_2 + {1 \over 2} M_N^{ij} N^c_i N^c_j \,,
\end{eqnarray}
where $N^c, E^c, L$ and $H_{1,2}$ are superfields of a singlet neutrino, charged lepton, $SU(2)_L$ lepton doublet and two Higgs doublet, respectively. 
The generations are denoted by $i$ and $j$.   
Yukawa matrices for charged leptons and neutrinos are defined as $Y_E$ and $Y_N$ respectively. 
A Majorana mass matrix is represented as $M_N$. 
The soft supersymmetry breaking terms in the lepton sector are given by\footnote{
We neglect the term $\tilde \nu_i^\dag \tilde \nu_j^\dag$~\cite{GOST}. 
}
\begin{eqnarray}
 - \mathcal L_\text{soft}^\text{lepton} 
 = (m_L^2)^{ij} \tilde \ell_i^\dag \tilde \ell_j +(m_E^2)^{ij} \tilde e_i^\dag \tilde e_j +(m_N^2)^{ij} \tilde \nu_i^\dag \tilde \nu_j 
    + ( T_E^{ij} \tilde e_i^\dag \tilde \ell_j h_1  +T_N^{ij} \tilde \nu_i \tilde \ell_j h_2 + \text{h.c.} ) \,,
\end{eqnarray}
where $\tilde f$ is a superpartner of $f$. 
The quark and gauge sector are defined in the same way as in the minimal supersymmetric standard model (MSSM). 
We follow the convention and notation defined by SUSY Les Houches Accord 2~\cite{SLHA} in the present paper.

At a low energy scale where the heavy fields $N^c_i$ are integrated out, the effective higher dimensional term is given as 
\begin{eqnarray}
 \label{Eq:SuperSeesaw}
 && W_\text{seesaw} =  {1 \over 2} K_N^{ij}  (L_i H_2) (L_j H_2) \,, \\
 \label{Eq:SSrelation}
 && K_N = -Y_N^T M_N^{-1} Y_N \,,
\end{eqnarray}
at the tree level. 
The neutrino mass matrix is obtained from this term after the electroweak symmetry is broken: 
\begin{eqnarray}
 \label{Eq:NuMass}
 m_\nu^{ij} = K_N^{ij} v^2 \sin^2 \beta \,,
\end{eqnarray}
where $v \simeq 174\, \text{GeV}$ and $\tan \beta$ is the ratio of two vacuum expectation values of the Higgs scalar fields in the superfields $H_1$ and $H_2$. 
Diagonalizing the mass matrix $m_\nu$ results in the tiny neutrino masses and the PMNS matrix.

We assume a universality of the soft SUSY breaking parameters as
\begin{eqnarray}
 \label{Eq:msugra}
 (m_L^2)^{ij} =(m_E^2)^{ij} =(m_N^2)^{ij} = M_0^2 \delta^{ij} \,,\quad T_N^{ij} = M_0 A_0 Y_N^{ij} \,,\quad T_E^{ij} = M_0 A_0 Y_E^{ij} \,,
\end{eqnarray}
at the GUT scale $\mu_G$, where $M_0$ is the universal scalar mass and $A_0$ is the dimensionless universal trilinear coupling. 
The soft breaking parameters in the squark and Higgs sector are also taken universal. 
For the gaugino masses, we introduce $M_{1/2}$ assuming the GUT relation. 
This ansatz clearly implies that the source of LFVs does not exist in the soft supersymmetry breaking terms at this scale of the Lagrangian while it does in the superpotential. 
For details on these assumptions and references, see in Ref.~\cite{GOST}. 

Below the GUT scale, however, the renormalization group equations (RGEs) of the parameters in Eq.~(\ref{Eq:msugra}) generate a slepton flavor mixing. 
A main source of the mixing is off-diagonal elements of $(Y_N^\dag Y_N)^{ij}$.  
In the approximation that all the singlet neutrinos are decoupled at a scale $\mu_R$, the contribution to the slepton mixing is represented as
\begin{eqnarray}
  \label{Eq:ML} (m_L^2)^{ij} &\simeq& -\frac{1}{8\pi^2} M_0^2 (3 + |A_0|^2) (Y_N^\dag Y_N)^{ij} \ln \frac{\mu_\text{G}}{\mu_R} \,, \\
  \label{Eq:ME}  (m_E^2)^{ij} &\simeq& 0 \,, \\
  \label{Eq:AE}  (T_E)^{ij} &\simeq& -\frac{1}{8\pi^2} M_0 A_0 \hat Y_E^{ii} (Y_N^\dag Y_N)^{ij} \ln \frac{\mu_\text{G}}{\mu_R} \,,
\end{eqnarray}
for $i \neq j$, where $\hat Y_E$ is the real positive matrix obtained by diagonalizing $Y_E$. 
To be more precise, we need to take into account the threshold effect because three right-handed neutrinos decouple at different mass scales.  
The precise treatment of this threshold effect modifies the calculation of the flavor mixing in the slepton sector. 
As explained in Sec.~\ref{Am}, we evaluate the LFVs by taking these effects.

\subsection{Structure of the neutrino Yukawa matrix}
Patterns of LFVs are considerably affected by the structure of matrices $Y_N$ and $M_N$. 
Here we summarize the parametrizations of them in our analysis. 
The superpotential for the lepton sector is given in Eq.~(\ref{Eq:SuperLepton}).
We decompose $Y_E, Y_N$ and $M_N$ in Eq.~(\ref{Eq:SuperLepton}) as
\begin{eqnarray}
  &&  Y_E =U_E^{[e]\dag} \hat Y_E U_L^{[e]}, \\
  &&  Y_N =U_N^{[\nu]\dag} \hat Y_N U_L^{[\nu]}, \\
  &&  M_N =U_N^{[M]\dag} \hat M_N U_N^{[M]*},
\end{eqnarray}
where $\hat Y_E, \hat Y_N$ and $\hat M_N$ are real positive diagonal matrices  
and $U_E^{[e]\dag}, U_N^{[\nu]\dag}, U_N^{[M]\dag}, U_L^{[e]}, U_L^{[\nu]}$ and $U_N^{[M]}$ are unitary matrices.  
We define the rotated fields as  
$L^{[a]} = U_L^{[a]} L$ (for $a=e,\nu$), $E^{c[e]} = U_{E}^{[e]*} E^c$ and $N^{c[b]} = U_{N}^{[b]*} N^c$ (for $b=\nu, M$). 
The superpotential in Eq.~(\ref{Eq:SuperLepton}) is written in terms of $E^{c[e]}, L^{[e]}$ and $N^{c[M]}$ as
\begin{eqnarray}
 W_\text{lepton} = 
 (W_\nu^T \hat Y_N V_\nu)^{ij} N^{c[M]}_i L^{[e]}_j H_2 
 + \hat Y_E^{ii} E^{c[e]}_i L^{[e]}_i H_1 + {1 \over 2} \hat M_N^{ii} N^{c[M]}_i N^{c[M]}_i \,,
\end{eqnarray}
where $V_\nu = U^{[\nu]}_N U^{[e]\dag}_L$ has three angles and one phase similarly to the Cabibbo-Kobayashi-Maskawa (CKM) matrix in the quark sector~\cite{CKM}, 
{\it i.e.}, $V_\nu$ is written as  
\begin{eqnarray}
 \label{Eq:Vnudef}
 V_\nu = 
 \begin{pmatrix} \bar c_{12} \bar c_{13} & \bar s_{12} \bar c_{13} & \bar s_{13} e^{-i\bar\delta_\nu} \\
 - \bar s_{12} \bar c_{23} - \bar c_{12} \bar s_{23} \bar s_{13} e^{i\bar\delta_\nu} & \bar c_{12} \bar c_{23} -\bar s_{12} \bar s_{23} \bar s_{13} e^{i\delta_\nu} & \bar s_{23} \bar c_{13} \\
\bar s_{12} \bar s_{23} -\bar c_{12} \bar c_{23} \bar s_{13} e^{i\bar\delta_\nu} & -\bar c_{12} \bar s_{23} -\bar s_{12} \bar c_{23} \bar s_{13} e^{i\delta_\nu} & \bar c_{23} \bar c_{13}  \end{pmatrix} ,\,
\end{eqnarray}
where $\bar c_{ij} =\cos\bar\theta_{ij}$, $\bar s_{ij} =\sin\bar\theta_{ij}$ with $0\leq\bar\theta_{ij}\leq\pi/2$ and $\bar\delta_\nu$ is a Dirac CP violating phase. 
The matrix $W_\nu = U^{[\nu]\dag}_N U^{[M]}_N$ is a special unitary matrix, which has three angles and five phases. 
Therefore, there are in total 18 free parameters in $V_\nu,W_\nu,\hat Y_N$ and $\hat M_N$, which cannot be reduced by redefinition of the fields. 
We define $Y_N^{[M,e]} = W_\nu^T \hat Y_N V_\nu$ and $Y_N^{[\nu,e]} = \hat Y_N V_\nu$ for later convenience.

On the other hand, the effective superpotential at the low energy scale is written as in Eq.~(\ref{Eq:SuperSeesaw}). 
We introduce another basis as $L^{[e']}= U_L^{[e']} L,\, E^{c[e']} =U_E^{[e']*} E^c$ for the charged leptons and $L^{[\nu']}= U_L^{[\nu']} L$ for the neutrinos 
so that $Y_E, K_N$ in Eq.~(\ref{Eq:SuperSeesaw}) are decomposed as
\begin{eqnarray}
 Y_E =U_E^{[e']\dag} \hat Y_E U_L^{[e']}, \quad K_N =U_L^{[\nu']T} \hat K_N U_L^{[\nu']}. 
\end{eqnarray}
In this case we can write the superpotential as, 
\begin{eqnarray}
 W_\text{lepton}^\text{eff} 
 = \hat Y_E^{ii} E^{c[e']}_i L^{[e']}_i H_1 + {1 \over 2} ( U_\nu^* \hat K_N U_\nu^\dag )^{ij}  (L_i^{[e']} H_2) (L_j^{[e']} H_2) \,,
\end{eqnarray}
where $U_\nu = U^{[e']}_L U^{[\nu']\dag}_L$ is the PMNS matrix which has three angles and three phases, {\it i.e.}, $U_\nu$ is defined as
\begin{eqnarray}
 \label{Eq:PMNSdef}
 U_\nu = 
 \begin{pmatrix} c_{12} c_{13} & s_{12} c_{13} & s_{13} e^{-i\delta_\nu} \\
 -s_{12} c_{23} -c_{12} s_{23} s_{13} e^{i\delta_\nu} & c_{12} c_{23} -s_{12} s_{23} s_{13} e^{i\delta_\nu} & s_{23} c_{13} \\
 s_{12} s_{23} -c_{12} c_{23} s_{13} e^{i\delta_\nu} & -c_{12} s_{23} -s_{12} c_{23} s_{13} e^{i\delta_\nu} & c_{23} c_{13}  \end{pmatrix} \,
 \begin{pmatrix} 1 & & \\ & e^{i\alpha_\nu/2} & \\ & & e^{i\beta_\nu/2} \end{pmatrix} \,,
\end{eqnarray}
where $c_{ij} =\cos\theta_{ij}$, $s_{ij} =\sin\theta_{ij}$ with $0\leq\theta_{ij}\leq\pi/2$, $\delta_\nu$ is a Dirac CP violating phase, 
and $\alpha_\nu,\beta_\nu$ are two Majorana CP violating phases. 
Neutrino masses are represented by $\hat K_N\, v^2 \sin^2\beta$. 
Therefore, 9 of 18 parameters in $V_\nu,W_\nu,\hat Y_N$ and $\hat M_N$ are used in order to generate three neutrino masses and the components of the PMNS matrix.
In other words, the LFV processes depend also on remaining 9 unfixed parameters.  
We note that the bases $[e]$ and $[e']$ are related as $L^{[e]}_i = P_L^{ii} L^{[e']}_i$ where $P_L^{ii}$ is a diagonal phase matrix. 
These two bases are different ($P_L \neq {\bf 1}$) in general with the phase conventions for $V_\nu$ and $U_\nu$ in Eqs.~(\ref{Eq:Vnudef}) and (\ref{Eq:PMNSdef}) respectively.

There are several choices for parametrizing $Y_N$ and $M_N$. 
In this paper, we use the following two parameterizations. 
\begin{itemize}
\item 
{\bf Parametrization 1}: We take the basis $Y_N^{[M,e']} =Y_N^{[M,e]} P_L=W_\nu^T \hat Y_N V_\nu P_L$. 
In this case, the Majorana mass matrix is diagonal and the neutrino Yukawa matrix is written as~\cite{IbarraLoad}   
\begin{eqnarray}
 \label{Eq:DegeCase}
 Y_N^{[M,e']} = \sqrt{\hat M_N} \,O_N \sqrt{\hat K_N} \,U_\nu^\dag \,,
\end{eqnarray}
where $O_N$ is a complex orthogonal matrix which is given by 
\begin{eqnarray}
 \label{Eq:Ortho}
 O_N = \hat M_N^{-1/2} W_\nu^T \hat Y_N V_\nu P_L U_\nu \hat K_N^{-1/2} \,.
\end{eqnarray} 
We have three parameters in $\hat M_N$ and six parameters in $O_N$. 
This parametrization befits a {\it degenerate} structure of $M_N$, since we can take $\hat M_N$ as input parameters. 

\item
{\bf Parametrization 2}: Another parametrization is to take the basis $Y_N^{[\nu,e']} =Y_N^{[\nu,e]} P_L = \hat Y_N V_\nu P_L$. 
In this basis, the Majorana mass matrix is written as 
\begin{eqnarray}
 \label{Eq:NonDcase}
 M_N^{[\nu]} =Y_N^{[\nu,e']} \left( U_\nu \hat K_N^{-1} U_\nu^T \right) Y_N^{[\nu,e']T} = W_\nu^* \hat M_N W_\nu^\dag .
\end{eqnarray}
The neutrino Yukawa matrix $Y_N^{[\nu,e']}$ contains 9 free parameters, and thus they control the contributions to LFVs via $Y_N^\dag Y_N$ as in Eqs.(\ref{Eq:ML})--(\ref{Eq:AE}). 
As will be explained later, it is convenient to apply this parametrization for a {\it non-degenerate} structure in $M_N$.
\end{itemize}

\section{Analysis method}
\label{Am}
In order to analyze low energy LFV signals in the supersymmetric seesaw model, 
it is required to evaluate the running effect on the parameters from the GUT scale $(\mu_\text{G})$ to the electroweak scale. 
At the GUT scale, we define the parameters of the soft breaking terms in the context of the minimal supergravity, that is $A_0,\,M_0$ and $M_{1/2}$. 
The neutrino Yukawa matrix $Y_N$ and the Majorana mass matrix $M_N$ are also defined at the GUT scale. 
After diagonalizing $M_N$, we obtain the masses of the right-handed neutrinos at their proper scales. 
Below the scales of right-handed neutrino masses, the soft breaking parameters are evaluated 
at the SUSY breaking and electroweak symmetry breaking (EWSB) scale $(\mu_\text{EWSB})$ by solving the RGEs.
Then, the physical mass spectrum and the flavor mixing matrices of SUSY particles are obtained at the electroweak scale. 
The detailed setup for the evaluation is summarized below.

\subsection{Neutrino sector}
Here we show the setup of the neutrino sector. 
The neutrino mass matrix $m_{\nu}^{ij}$ obtained at the low energy scale is decomposed as 
\begin{eqnarray}
 m_{\nu}^{ij} = (U_\nu^*)^{ik} m_{\nu_k}  (U_\nu^\dag)^{kj} \,,
\end{eqnarray}
where $U_\nu$ is the PMNS matrix defined in Eq.~(\ref{Eq:PMNSdef}) and $m_{\nu_k}$ is a neutrino mass eigenvalue. 
Since the two squared mass differences of the neutrinos $\Delta m^2_{ij} = m_{\nu_i}^2 -m_{\nu_j}^2$ satisfy $|\Delta m_{32}^2| \gg |\Delta m_{21}^2|$, 
the neutrino mass spectra can be hierarchical. 
The cases for $m_{\nu_3} \gg m_{\nu_2} > m_{\nu_1}$ and $m_{\nu_2} > m_{\nu_1} \gg m_{\nu_3}$ are referred as normal and inverted hierarchies, respectively. 
In our analysis, we consider both cases.

\subsection{Renormalization group equations}
We solve the RGEs of the SUSY parameters including the seesaw sector by using the public code {\it SPheno 3.2.4} written by W.~Porod and F.~Staub \cite{Porod}.
Two loop running effects and complete one loop corrections to all SUSY and Higgs particle masses are included as explained in Ref.~\cite{Porod}. 
As for the Higgs boson, its pole mass is calculated at the two loop level. 
The setup in our study is listed as follows: 
\begin{itemize}
 \item The GUT scale is set to be $\mu_\text{G} = 2\times 10^{16}\, \text{GeV}$.
 \item The EWSB scale is determined at $\mu_\text{EWSB} = \sqrt{m_{\tilde t_1} m_{\tilde t_2}}$, where $m_{\tilde t_{1,2}}$ are stop masses.
 \item The neutrino Yukawa matrix with a specific structure is defined at the GUT scale.
 \item The higgsino mass parameter $\mu$ and $A_0$ are assumed to be real to avoid constraints from experimental searches for various electric dipole moments~\cite{Ellis:1982tk}. 
\end{itemize}

We take into account the threshold effect mentioned in Sec.~\ref{SubSec:seesaw model} by integrating out right-handed neutrinos one by one at their proper scales. 
This effect generates contributions to the slepton mixing in addition to those given in Eqs.~(\ref{Eq:ML})-(\ref{Eq:AE}). 
Furthermore, the seesaw relation shown in Eq.~(\ref{Eq:SSrelation}) is modified~\cite{REAP,EllisRaidal}.

Among 18 parameters in $Y_N$ and $M_N$, we choose 9 of them as inputs at the GUT scale and adjust the others to reproduce the neutrino masses $m_{\nu_i}$ and the PMNS matrix $U_\nu$. 
To do that, we define $m_{\nu_i}^\text{G}$ and $U_\nu^\text{G}$ at the GUT scale as
\begin{eqnarray}\label{Eq:seesawGUT}
  (U_\nu^{\text{G}*})^{ik}  m_{\nu_k}^\text{G}  (U_\nu^{\text{G}\dag})^{kj} = -v^2 \sin^2 \beta\, (Y_N^T M_N^{-1} Y_N)^{ij} \,. 
\end{eqnarray}
We numerically determine $U_\nu^\text{G}$ and $m_{\nu}^\text{G}$ which reproduce the PMNS matrix and the neutrino masses at the low energy scale by an iterative method.  
We note that $s_{12}^\text{G}$, $\Delta m_{32}^{2,\text{G}}$ and $\Delta m_{21}^{2,\text{G}}$ in Eq.~(\ref{Eq:seesawGUT}) are sensitive 
to the running effect compared with the other components~\cite{REAP,EllisRaidal}.

\subsection{Structure of $Y_N$}\label{SubSec:YN}
In the present work, we investigate LFV signals in two cases with specific structures in $Y_N$ and $M_N$: {\it degenerate case} and {\it non-degenerate case}. 
For each case, we apply the appropriate parametrization which we have shown in the previous section.

\begin{flushleft}
{\it Degenerate case}: 
\end{flushleft}
First, we consider {\it degenerate case} (D case), which means that the Majorana mass matrix is assumed to be proportional to the unit matrix. 
In this case, we apply Parametrization 1.   
We decompose the matrix $O_N$ as 
\begin{eqnarray}
 O_N = \tilde O_N e^{i A_N}\,, \quad
 A_N = \begin{pmatrix} 0&a&b \\-a&0&c \\-b&-c&0  \end{pmatrix} \label{Eq:ANdef}
\end{eqnarray}
where $\tilde O_N$ is a real orthogonal matrix and $A_N$ is a real anti-symmetric matrix $A^T_N =-A_N$. 
The matrix $\tilde O_N$ is irrelevant for the LFV signals since the source of the flavor mixing comes from $Y_N^\dag Y_N$.   
Thus we take $\tilde O_N ={\bf 1} $ without loss of generalities. 
The neutrino Yukawa matrix $Y_N$ is written as 
\begin{eqnarray}
 \label{Eq:DcaseSt}
 Y_N = \frac{\sqrt{\tilde M_N}}{v\sin\beta}\,\, e^{iA_N} \begin{pmatrix} \sqrt{m_{\nu_1}} & & \\ & \sqrt{m_{\nu_2}} & \\ & &\sqrt{m_{\nu_3}} \end{pmatrix} U_\nu^\dag \,,
\end{eqnarray}
where $\tilde M_N$ is the degenerate Majorana mass eigenvalue. 
As for the parameters in $A_N$, we take $a=b=0\, (b=c=0)$ for normal (inverted) hierarchical mass spectrum of the neutrinos 
since the contributions of $a$ and $b$ ($b$ and $c$) are subdominant according to the analysis in Ref.~\cite{ShinPetc}. 
This can be understood as follows. 
If we expand the off-diagonal element $(Y_N^\dag Y_N)_{ij}$ by $a,b$ and $c$ assuming $|a|,|b|,|c| \ll 1$, 
the contributions of $a,b$ and $c$ appear at the leading order in the combinations of $|a| \sqrt{m_{\nu_1}m_{\nu_2}},\,|b| \sqrt{m_{\nu_1}m_{\nu_3}}$ and $|c| \sqrt{m_{\nu_2}m_{\nu_3}}$. 
In the case of the normal hierarchy, the contributions of $|a| \sqrt{m_{\nu_1}m_{\nu_2}}$ and $|b| \sqrt{m_{\nu_1}m_{\nu_3}}$ are not significant 
compared with $|c| \sqrt{m_{\nu_2}m_{\nu_3}}$. 
For example, the element $(Y_N^\dag Y_N)_{12}$ which induces $\mu\to e\gamma$ can be represented as 
\begin{align}
 (Y_N^\dag Y_N)_{12} \simeq \frac{\tilde M_N}{v^2 \sin^2\beta} 
 &\Bigg( \left( m_{\nu_2} -m_{\nu_1} \right) \,c_{12} s_{12} c_{23} +\left( m_{\nu_3} -m_{\nu_2} \right)  \,s_{13} s_{23} e^{-i\delta_\nu} \notag \\
 &+2i\, c \sqrt{ m_{\nu_2} m_{\nu_3} } \,s_{12} s_{23} e^{i(\alpha_\nu-\beta_\nu)} \Bigg), \label{Eq:YNYN12}
\end{align}
with this approximation. 
The expression for the inverted hierarchy is obtained in a similar way.

\begin{flushleft}
{\it Non-degenerate case}: 
\end{flushleft}
Second, we consider {\it non-degenerate case} (ND case) for the structure of $M_N$. 
In this case, we apply Parametrization 2 and $Y_N^\dag Y_N$ can be considered as an input. 
Thus the Majorana mass matrix $M_N$ is determined by Eq.~(\ref{Eq:NonDcase}). 
To see how large $\mathcal B(\tau\to\mu\gamma)$ can be within the constraint on $\mathcal B(\mu \to e\gamma)$, we take $\bar\theta_{12}=\bar\theta_{13}=0$.  
Accordingly, $Y_N$ is parametrized as 
\begin{eqnarray}
\label{Eq:NDcaseSt}
 Y_N = 
 \begin{pmatrix} y_{1} & & \\ & y_{2} & \\ & & y_{3} \end{pmatrix} 
 \begin{pmatrix} 1 & 0 & 0 \\  0 & \cos \theta & \sin \theta \\ 0 & -\sin \theta & \cos \theta \end{pmatrix} P_L \,,
\end{eqnarray}
where $\theta = \bar\theta_{23}$. 
We also take $P_L = {\bf 1}$ for simplicity and consider $y_1,y_2,y_3 \lesssim O(1)$. 
Even though $\mu \to e \gamma$ and $\tau \to e \gamma$ do not occur in the approximation using Eqs.~(\ref{Eq:ML})--(\ref{Eq:AE}), 
the threshold effects generate non zero contributions to $\mu \to e \gamma$ and $\tau \to e \gamma$.  
In the case of $\bar\theta_{12}=\bar\theta_{23}=0$, a similar consideration can be applied for $\tau\to e\gamma$.

\subsection{Observables}
In this subsection, we summarize the formulae of relevant processes. 
The LFV process emitting a photon, $\ell_j \to \ell_i\gamma$, is generated by $\mathcal O_7$ operator which is defined as 
\begin{eqnarray}
&& \mathcal L_\text{eff}^{\ell \to \ell' \gamma} =C_{7L}^{ij} \mathcal O_{7L}^{ij} +C_{7R}^{ij} \mathcal O_{7R}^{ij} +\text{h.c.} \,, \\
&& \mathcal O_{7^L_R}^{ij} = F_{\mu\nu}\, \bar\ell_i \sigma^{\mu\nu} \left( \frac{1\mp\gamma^5}{2} \right) \ell_j \,,
\end{eqnarray}
where $\sigma^{\mu\nu} = (i/2) [\gamma^\mu, \gamma^\nu]$ and the coefficients $C_{7L,\,7R}^{ij}$ are obtained from contributions via lepton flavor mixing loop diagrams. 
Then, the decay rate is given by 
\begin{eqnarray}
 \Gamma (\ell_j \to \ell_i \gamma) = \frac{m_{\ell_j}^3}{4\pi} \left(  |C_{7L}^{ij}|^2 +|C_{7R}^{ij}|^2 \right)  \,, 
\end{eqnarray}
where we neglect the lepton mass in the final state. 
In the SM, neutrino mixings contribute to $C_{7R}^{ij}$ with a strong suppression factor as $\Delta m_{ij}^2/m_W^2$. 
In a supersymmetric model, since sleptons and sneutrinos also carry flavor indices, 
loop diagrams with sleptons or sneutrinos affect $\ell_j \to \ell_i \gamma$. 
The coefficients $C_{7L,\,7R}^{ij}$ are written in terms of masses and flavor mixing matrices of SUSY particles~\cite{LFVformulaPorod,LFVformulaHisano}.

For the quark sector, extra flavor mixings exist in a supersymmetric model.  
Even if the squark mass matrix is assumed to be diagonal at the GUT scale, off-diagonal elements are generated by the RGE. 
The off-diagonal elements induce the SUSY contributions to flavor changing observables via loop diagrams in the quark sector. 
Among them, $b \to s$ transition processes such as $B \to X_s \gamma$ and $B_s \to \ell^+\ell^-$ are important. 
The effective Lagrangians for these processes are given by    
\begin{eqnarray}
&&	\mathcal L_\text{eff}^{B_s \to \ell^+\ell^-} = 
	\frac{\alpha G_F}{ \sqrt2 \pi \sin^2 \theta_W} V_{ts}^*V_{tb} \sum_{i=A,S,P} \left( C_i \mathcal O_i + C'_i \mathcal O'_i \right) +\text{h.c.} \,, \\
&&	\mathcal O_A = \left(\bar s_L \gamma_\mu b_L \right) \left( \bar \ell \gamma^\mu \gamma_5 \ell \right) \,,\quad 
	\mathcal O'_A = \left(\bar s_R \gamma_\mu b_R \right) \left( \bar \ell \gamma^\mu \gamma_5 \ell \right) \,, \\
&&	\mathcal O_S =m_b \left(\bar s_R b_L \right) \left( \bar \ell \ell \right) \,,\quad 
	\mathcal O'_S =m_s \left(\bar s_L b_R \right) \left( \bar \ell \ell \right) \,, \\
&&	\mathcal O_P =m_b \left(\bar s_R b_L \right) \left( \bar \ell \gamma^5 \ell \right) \,,\quad 
	\mathcal O'_P =m_s \left(\bar s_L b_R \right) \left( \bar \ell \gamma^5 \ell \right) \,, 
\end{eqnarray}
and
\begin{eqnarray}
&&	\mathcal L_\text{eff}^{B \to X_s \gamma} = 
	2\sqrt 2 G_F V_{ts}^*V_{tb} \sum_{i=7,8} \left( C_i \mathcal O_i + C'_i \mathcal O'_i \right) +\text{h.c.} \,, \\
&&	\mathcal O_7 =\frac{e}{16\pi^2}m_b\bar s_L \sigma^{\mu\nu} b_R F_{\mu\nu} \,,\quad \mathcal O'_7 = \frac{e}{16\pi^2} m_s \bar s_R \sigma^{\mu\nu} b_L F_{\mu\nu} \,, \\
&&	\mathcal O_8 =\frac{g_s}{16\pi^2}m_b\bar s_L \sigma^{\mu\nu}T^a b_R G^a_{\mu\nu} \,,\quad \mathcal O'_8 = \frac{g_s}{16\pi^2} m_s \bar s_R \sigma^{\mu\nu} T^a b_L G^a_{\mu\nu}\,.
\end{eqnarray}
The contributions of SUSY particles are all included in the Wilson coefficients $C_i^{(\prime)}$. 
The analytical formulae for $\mathcal B (B_s \to \ell^+\ell^-)$ and $\mathcal B (B \to X_s \gamma)$ are found 
in Refs.~\cite{BSformulaBuras} and \cite{BSformulaLunghi}, respectively.

In the supersymmetric model, the Higgs boson mass is less than the $Z$ boson mass at the tree level and increases owing to a radiative correction~\cite{HiggsMassLoop}. 
The Higgs boson mass is evaluated including two-loop corrections following the formula in Ref.~\cite{HiggsformulaDedes}, which is implemented in {\it SPheno}.

\section{Numerical result}
\label{Nr}
In this section, we present the allowed region of the SUSY parameter space and predictions on the patterns of the LFV signals.

\subsection{Inputs and constraints}
For the neutrino parameters, we adjust the parameters at the GUT scale so that the mass differences and mixings are consistent with the present neutrino oscillation data~\cite{PDGneutrino}. 
To do that, we apply the allowed range for the neutrino parameters at the low energy scale shown in Table~\ref{Tab:NuInputs}. 
In our calculation, we take the lightest neutrino mass to be $m_{\nu_{1(3)}} = 0.0029$ - $0.0031\,\text{eV}$ in the normal (inverted) hierarchy.  
\begin{table}\begin{center}\begin{tabular}{c|c|c}
 \hline 
 \hline Measurement 								& Experimental result~\cite{PDGneutrino}				& Allowed range				 \\
 \hline $\Delta m_{21}^2 \times 10^5 \,\text{eV}^{-2}$	& $7.54^{+0.26}_{-0.22} \,\,(7.54^{+0.26}_{-0.22})$ 		& $7.3$\,--\,$7.8$			 	 \\
 \hline $|\Delta m_{32}^2| \times 10^3 \,\text{eV}^{-2}$	& $2.39\pm 0.06 \,\,(2.42\pm 0.06)$						& $2.3$\,--\,$2.5$ 				 \\
 \hline $\sin^2 \theta_{12}$							& $0.308\pm 0.017 \,\, (0.308\pm 0.017)$ 				& $0.29$\,--\,$0.33$	 			 \\
 \hline $\sin^2 \theta_{23}$							& $0.437^{+0.033}_{-0.023} \,\, (0.455^{+0.039}_{-0.031})$	& $0.41$\,--\,$0.49$				 \\
 \hline $\sin^2 \theta_{13}$							& $0.0234^{+0.0020}_{-0.0019} \,\, (0.0240^{+0.0019}_{-0.0022})$	& $0.022$\,--\,$0.026$ 		\\
 \hline\hline
\end{tabular}\end{center}
\caption{
Experimental results and allowed ranges for the neutrino parameters which we take into account in our numerical calculation.
The experimental values are results from the fitted analysis assuming the normal (inverted) hierarchy of neutrino masses. 
We apply the same allowed ranges for both normal and inverted hierarchical case.   
}\label{Tab:NuInputs}
\end{table}
We emphasize that the angle $\theta_{13}$ has been precisely determined by experiments and found to be not zero~\cite{Theta13T2K,Theta13MINOS,Theta13Chooz,Theta13Daya,Theta13RENO}. 
By the combination of the reactor experiments and the experiment of electron neutrino appearance from a muon neutrino beam in Refs.~\cite{DeltaCPMINOS,DeltaCPT2K}, 
a preferred range of $\delta_\nu$ can be obtained but the constraint is not strong. 
As for the Majorana CP violating phases, there is no experimental constraint. 
In our analysis, we treat the CP violating phases as free parameters.

Recent results at LHC experiments put constraints on the SUSY parameter space. 
First, the Higgs boson mass is measured as $m_h \simeq 126\, \text{GeV}$~\cite{ATLAShiggs, CMShiggs}.
For the MSSM, this value implies a heavy stop or a large left-right mixing in the stop sector. 
Second, direct searches for the squarks and the gluino impose the constraints on their masses~\cite{ATLASsusy, CMSsusy}. 
The lower limit of the gluino mass is around $1.4 \,\text{TeV}$ and that of the squark masses for the first and second generations is about $1.7 \,\text{TeV}$.

Flavor experiments have also improved their results over the past years. 
It is known that the branching ratio of $\bar B\to X_s\gamma$ put a severe constraint on the SUSY parameter space. 
Including the recent updated result of the {\sc BaBar} experiment~\cite{bsgBABAR2012}, 
the latest world average of the branching ratio is $\mathcal B(\bar B\to X_s\gamma) = (3.43 \pm 0.21 \pm 0.07) \times 10^{-4}$ \cite{bsgHFAG}. 
Another important process is $B_s \to \mu^+\mu^-$, which was first observed by the LHCb collaboration \cite{BsMuMuLHCbFirst}. 
The latest result is $\mathcal B(B_s \to \mu^+\mu^-) = (2.9 \pm 0.7) \times 10^{-9}$ \cite{BsMuMuComb} 
combining the results obtained by the LHCb \cite{BsMuMuLHCb} and CMS \cite{BsMuMuCMS} collaborations.

For LFV processes, an improved constraint on $\mu\to e\gamma$ is obtained as $\mathcal B(\mu\to e\gamma) < 5.7 \times 10^{-13}$ 
at $90\%$ confidence level (CL) by the MEG collaboration~\cite{MuEgMEG}. 
The flavor violating decays of the tau lepton, $\tau\to\mu\gamma$ and $\tau\to e\gamma$, are also constrained by the {\sc BaBar}~\cite{TauMuEbabar} and Belle~\cite{TauMuEbelle} collaborations.
The recent bounds on the branching ratios are $\mathcal B(\tau\to\mu\gamma) < 4.4 \times 10^{-8}$ and $\mathcal B(\tau\to e\gamma) < 3.3 \times 10^{-8}$.

\begin{table}\begin{center}\begin{tabular}{c|c|c}
 \hline 
 \hline Measurement 									& Experimental result										& Allowed range						 \\
 \hline $m_h$											
 & \begin{tabular}{c} ATLAS: $(125.36 \pm 0.37 \pm 0.18 ) \,\text{GeV}$ \cite{ATLAShiggsRecent} 
 \\ CMS: $(125.03^{+0.26}_{-0.27} \,^{+0.13}_{-0.15} ) \,\text{GeV}$ \cite{CMShiggsRecent}\end{tabular}
 & $(126\pm 2)\text{GeV}$			 	 \\
 \hline $\mathcal B(\bar B\to X_s\gamma) \times 10^4$	& $3.43 \pm 0.21 \pm 0.07$ \cite{bsgHFAG}					& $3.43\pm 0.62$ 					 \\
 \hline $\mathcal B(B_s \to \mu^+\mu^-) \times 10^9$		& $2.9 \pm 0.7$ \cite{BsMuMuComb}						& $2.9\pm 0.7$					        \\
 \hline $\mathcal B(\mu \to e\gamma) \times 10^{13}$		& $<5.7$ \cite{MuEgMEG}									& $<5.7$  							 \\
 \hline $\mathcal B(\tau\to\mu (e) \gamma) \times 10^8$	& $<4.4(3.3)$ \cite{TauMuEbabar}							& $<4.4(3.3)$ 						 \\
 \hline $M_{\tilde g}$									& $>1.4$ TeV \cite{ATLASsusy, CMSsusy}					& $>1.4$ TeV 						 \\
 \hline $M_{\tilde q}$									& $>1.7$ TeV \cite{ATLASsusy, CMSsusy}					& $>1.7$ TeV					 	 \\
 \hline\hline
\end{tabular}\end{center}
\caption{Experimental results on flavor signals and masses and their allowed ranges taken in our analysis.}\label{Tab:Bounds}
\end{table}
In order to take these constraints into account, we apply the allowed ranges as listed in Table~\ref{Tab:Bounds}. 
For the allowed range of $\mathcal B(\bar B\to X_s\gamma)$, we include theoretical uncertainty~\cite{bsgThError} and experimental uncertainties within $2\sigma$ ranges.   
In order to take the theoretical uncertainties of the Higgs boson mass into account, we assign $\pm 2\,\text{GeV}$ for the allowed range as shown in Table~\ref{Tab:Bounds}. 
We take into account the only experimental error in $\mathcal B(B_s \to \mu^+\mu^-)$. 
As for the SUSY particle mass bounds, the quoted bounds are used to constrain the SUSY parameter space in our analysis, 
even though these bounds are obtained under some assumption for the SUSY spectrum.

\begin{figure}[!ht]
\includegraphics[viewport=0 0 650 510,width=42em]{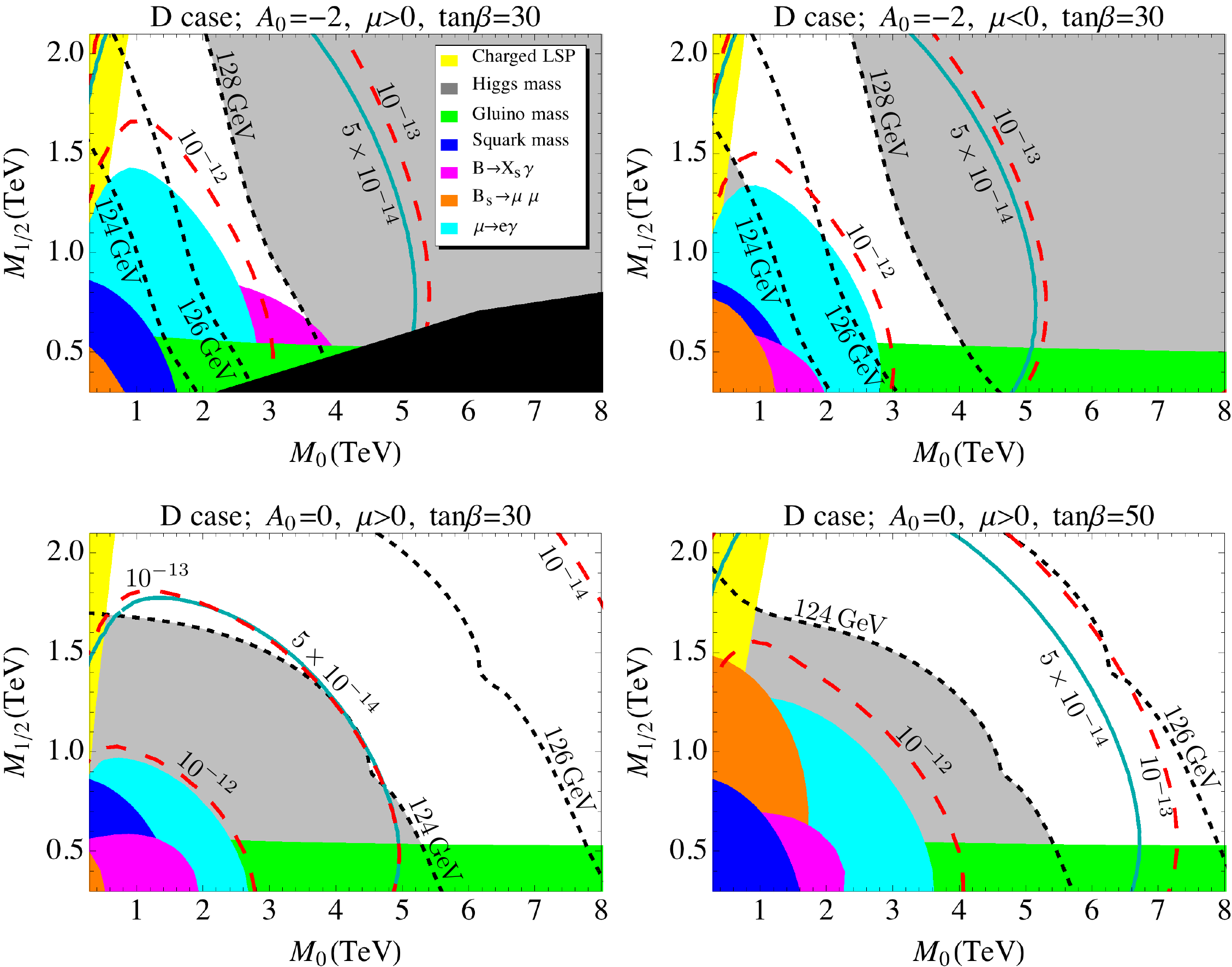} 
\caption{
Excluded region plots for $(M_0,\,M_{1/2})$ with fixed parameters of $(\text{sign}(\mu),\,A_0,\,\tan\beta)$ in the degenerate case with the normal hierarchical neutrino masses. 
The parameters in the neutrino sector are assumed as $\tilde M_N=7 \times 10^{12}\, \text{GeV}$ and $O_N={\bf 1}$. 
Each colored region is excluded by the observable as exhibited in the legend. 
The dotted, dashed and solid lines show the contours of $m_h$, $\mathcal B(\tau\to\mu\gamma)$ and $\mathcal B(\mu\to e\gamma)$, respectively. 
}
\label{FIG:constD}
\end{figure}
\begin{figure}[!ht]
\includegraphics[viewport=0 0 650 510,width=42em]{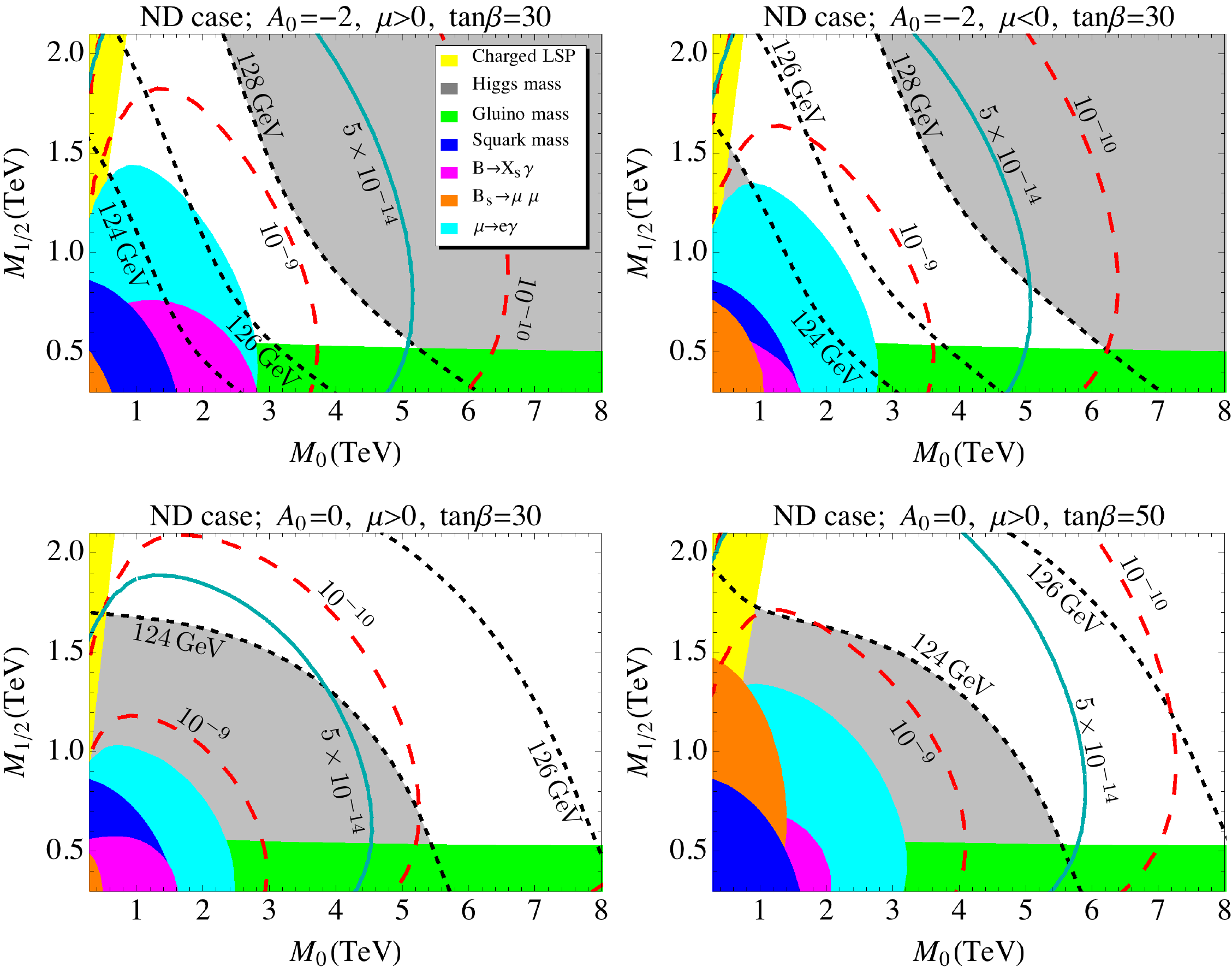} 
\caption{
Excluded region plots in the non-degenerate case with the normal hierarchical neutrino masses.  
The parameters in $Y_N$ are assumed as $y_1=0.05,\, y_2=y_3=1.5$ and $\theta=\pi/4$.  
The notation is the same as Fig.~\ref{FIG:constD}. 
}
\label{FIG:constND}
\end{figure}
In Figs.~\ref{FIG:constD} and \ref{FIG:constND}, we show the excluded region plots for the $(M_0,\,M_{1/2})$ plane by taking into account the constraints discussed above. 
In the figures, colored regions are excluded and the white regions are allowed. 
We take the normal hierarchy for neutrino masses and the CP violating phases in the PMNS matrix are assumed as $\delta_\nu = \alpha_\nu = \beta_\nu =0$. 
For the neutrino Yukawa matrix $Y_N$, we consider the degenerate and the non-degenerate cases in Figs.~\ref{FIG:constD} and \ref{FIG:constND}, respectively. 
For the degenerate case, we take $\tilde M_N=7 \times 10^{12}\,\text{GeV}$ and $O_N={\bf 1}$. 
For the non-degenerate case, we take $y_1=0.05,\, y_2=y_3=1.5$ and $\theta=\pi/4$. 
As for the SUSY parameters, $\text{sign}(\mu),\,A_0$ and $\tan\beta$ are fixed as shown in the figures. 
In the yellow region, the lightest SUSY particle (LSP) is a charged particle. 
In the black region in Fig.~\ref{FIG:constD}~(a), the EWSB does not occur. 
The black dotted lines show the contours of the Higgs boson mass for $m_h=124,\,126$ and $128\,\text{GeV}$,  
and the gray region shows that the Higgs boson mass is outside of the allowed range. 
The blue and green regions are excluded by the lower mass bounds of the squark and gluino, respectively. 
The magenta, orange and cyan regions are not allowed by the experimental data of $\mathcal B(\bar B\to X_s\gamma)$, $\mathcal B(B_s \to \mu^+\mu^-)$ and $\mathcal B(\mu \to e\gamma)$, respectively.
We also show the contours of $\mathcal B(\mu\to e\gamma)$ and $\mathcal B(\tau\to\mu\gamma)$ with the cyan solid and red dashed lines, respectively.  
As is well-known, the allowed region from the Higgs boson mass depends on $A_0$, since the left-right mixing in the stop sector affects the Higgs boson mass. 
One can see that the squark and slepton masses are large for $A_0=0$ and their masses are relatively small for $A_0=-2$.
Moreover the lepton flavor mixing is enhanced for $A_0=-2$. 
Thus the LFV decay rates are lager for $A_0=-2$ than those for $A_0=0$. 
It can be seen from Figs.~\ref{FIG:constD} and \ref{FIG:constND} in the same set up for $(\text{sign}(\mu) ,A_0, \tan\beta)$ that the Higgs boson mass is also affected by the structure in $Y_N$. 
This is because $Y_N$ alters the renormalization group running of the stop $A$ term and the stop masses. 
The constraints from the $B$ physics are also important. 
In a certain parameter region, only $\mathcal B(\bar B\to X_s\gamma)$ gives a constraint. 
In Fig.~\ref{FIG:constD}, only $\mathcal B(\mu\to e\gamma)$ can be large in the allowed region, 
whereas both $\mathcal B(\mu\to e\gamma)$ and $\mathcal B(\tau\to\mu\gamma)$ could be close to the current experimental upper bounds in the case of Fig.~\ref{FIG:constND}. 
We note that our result is consistent with the other recent studies in Refs.~\cite{Figueiredo:2013tea,Calibbi:2012gr}.
In the allowed regions in Figs.~\ref{FIG:constD} and \ref{FIG:constND}, the SUSY contribution to the muon $g-2$ is too small to explain the present deviation~\cite{MuonG}. 
In this study, we do not use the muon $g-2$ as a constraint.

\subsection{LFV signals}
We investigate signals of LFV for the degenerate case and the non-degenerate case in the neutrino sector.  
In particular, we discuss the correlation between $\mathcal B(\mu\to e\gamma)$ and $\mathcal B(\tau\to\mu\gamma)$.

\subsubsection{Degenerate case}
\begin{figure}
\includegraphics[viewport=0 0 730 400,width=36em]{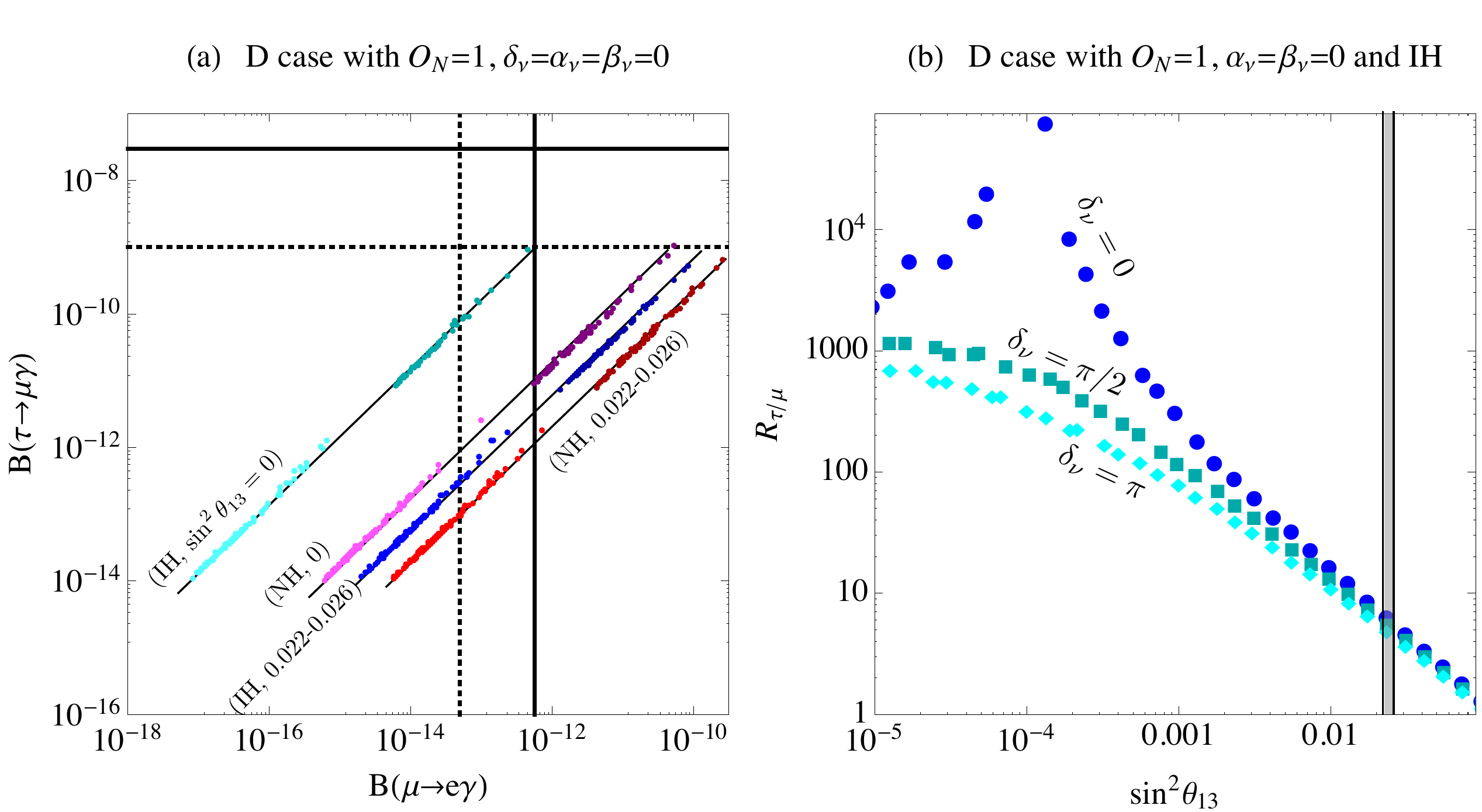} 
\caption{
(a) Correlation between $\mathcal B(\mu \to e\gamma)$ and $\mathcal B(\tau\to\mu\gamma)$ in the degenerate case with $O_N={\bf 1}$.  
The SUSY parameters $(M_{1/2},\,M_0,\,A_0)$ are randomly generated and we take $\tan\beta=30$ and $\mu>0$. 
The upper-right group of dots along each line corresponds to $\tilde M_N = 5\times 10^{14}\,\text{GeV}$ and the lower-left one corresponds to $\tilde M_N = 7\times 10^{12}\,\text{GeV}$. 
The results of NH and IH within $\sin^2 \theta_{13}=0.022$-$0.026$ and with $\sin^2 \theta_{13}=0$ are shown as indicated. 
The vertical and horizontal solid lines show the present upper bounds of $\mathcal B(\mu \to e\gamma)$ and $\mathcal B(\tau\to\mu\gamma)$, respectively. 
The dotted lines indicate expected sensitivities at SuperKEKB/Belle~II and MEG~II. 
(b) The ratio $R_{\tau/\mu}=\mathcal B(\tau\to\mu\gamma)/\mathcal B(\mu \to e\gamma)$ for IH as a function of $\sin^2 \theta_{13}$. 
The gray region is the present experimental value of $\sin^2 \theta_{13}$. 
}
\label{FIG:scattD}
\end{figure}
In Ref.~\cite{GOST}, it has been found that $\mathcal B(\mu\to e\gamma)$ and $\mathcal B(\tau\to e\gamma)$ can be suppressed while keeping a large $\mathcal B(\tau\to\mu\gamma)$ 
in the simplest degenerate case with $O_N = {\bf 1}$ in Eq.~(\ref{Eq:DcaseSt}), if $\theta_{13}$ is chosen to be zero and neutrino masses are inversely hierarchical. 
This is because the off-diagonal elements of $Y_N^\dag Y_N$ are approximately written as, 
\begin{eqnarray}
 (Y_N^\dag Y_N)_{12,\,13} \propto \frac{\tilde M_N}{v^2} \cdot \frac{\Delta m_{21}^2}{m_{\nu_1} + m_{\nu_2}}  \,,
 \quad 
 (Y_N^\dag Y_N)_{23} \propto \frac{\tilde M_N}{v^2} \cdot \frac{\Delta m_{32}^2}{m_{\nu_2} + m_{\nu_3}}  \,,
\end{eqnarray}
for $\theta_{13}=0$ in the PMNS matrix. 
Thus $\mathcal B(\mu\to e\gamma)$ and $\mathcal B(\tau\to e\gamma)$ are strongly suppressed for the inverted hierarchical case.

Taking into account recent experimental results from neutrino experiments, we show the correlation between $\mathcal B (\mu\to e\gamma)$ and $\mathcal B (\tau\to\mu\gamma)$ 
in the simplest degenerate case within $\sin^2 \theta_{13}=0.022$-$0.026$ in Fig.~\ref{FIG:scattD}~(a).
We also present the result with $\sin^2 \theta_{13}=0$ for comparison. 
We take $\delta_\nu = \alpha_\nu = \beta_\nu =0$ for both NH and IH cases. 
As for the SUSY inputs, we randomly vary them within 
\begin{eqnarray}
 0< M_0 < 8\,\text{TeV}\,, \quad 0< M_{1/2} < 2\,\text{TeV}\,, \quad -2<A_0<2\,,
\end{eqnarray}
and taking the others as $\mu>0$, $\tan\beta=30$.  
The upper-right group of dots along each line corresponds to the Majorana mass $\tilde M_N = 5\times 10^{14}\,\text{GeV}$ and 
the lower-left one corresponds to $\tilde M_N = 7\times 10^{12}\,\text{GeV}$. 
We plot points allowed by the constraints listed in Table~\ref{Tab:Bounds} except for $\mathcal B (\mu\to e\gamma)$ and $\mathcal B (\tau\to\mu\gamma)$. 
The vertical and horizontal solid lines represent the present experimental upper limits on $\mathcal B (\mu\to e\gamma)$ and $\mathcal B (\tau\to\mu\gamma)$, respectively. 
We also show the possible reaches of $\mathcal B (\tau\to\mu\gamma) \simeq 10^{-9}$ expected at SuperKEKB/Belle~II 
and $\mathcal B (\mu\to e\gamma) \simeq 5\times 10^{-14}$ at MEG~II~\cite{Baldini:2013ke} with dotted lines.  
We find that the ratio $R_{\tau/\mu} = \mathcal B(\tau\to\mu\gamma)/\mathcal B(\mu \to e\gamma)$ is insensitive to $\tilde M_N$ and the SUSY parameters, 
whereas it is sensitive to $\theta_{13}$ and the mass ordering. 
We also show $R_{\tau/\mu}$ as a function of $\sin^2 \theta_{13}$ for the inverted hierarchical cases with $\delta_\nu=0, \pi/2$ and $\pi$ in Fig.~\ref{FIG:scattD}~(b). 
Although $R_{\tau/\mu}$ is enhanced by several orders of magnitude if $\sin^2 \theta_{13} \sim 10^{-4}$, 
such an enhancement does not occur for the present value of $\sin^2 \theta_{13}$.

\begin{figure}[!ht]
\includegraphics[viewport=0 0 730 400,width=36em]{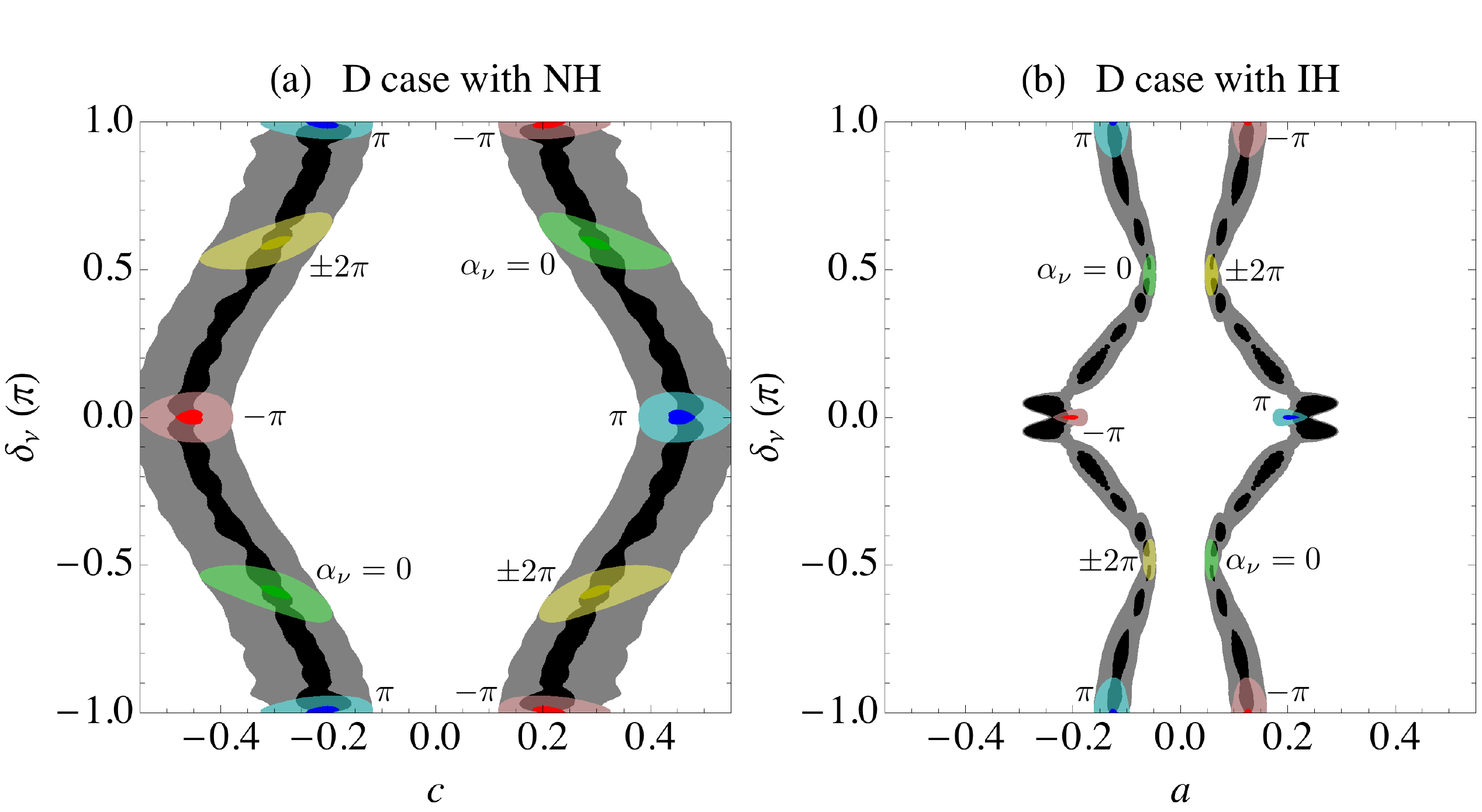} 
\caption{
Contour plots of $R_{\tau/\mu}$ on the planes of $(c, \delta_\nu)$ and $(a, \delta_\nu)$ for (a) normal and (b) inverted hierarchical cases. 
Black (gray) region corresponds to $R_{\tau/\mu} > 1800$ ($100$). 
The regions of $R_{\tau/\mu} > 1800$ ($100$) with the Majorana CP violating phases being fixed as $\beta_\nu=0,\, \alpha_\nu =0,\,\pm\pi$ and $\pm 2\pi$ are also shown with darker (lighter) color. 
}
\label{FIG:contD}
\end{figure}
The signals of LFV depend on the phase parameters $\delta_\nu$, $\alpha_\nu$ and $\beta_\nu$ in the PMNS matrix and $a$, $b$ and $c$ defined in Eq.~(\ref{Eq:ANdef}). 
In the normal hierarchical case, the Majorana CP violating phase contribution always appears in the form of $\alpha_\nu - \beta_\nu$ because $m_{\nu_1} \ll m_{\nu_2}$. 
As explained in the Sec.~\ref{SubSec:YN}, the contribution of $a$ and $b$ is negligible. 
Therefore we take $\beta_\nu=a=b=0$. 
Similarly, in the inverted hierarchical case, we take $\beta_\nu=b=c=0$.  
In Fig.~\ref{FIG:contD}, the region in which the maximal value of $R_{\tau/\mu}$ in the range $-2\pi \leq \alpha_\nu \leq 2\pi$ is larger than $1800$ ($100$) is shown in dark (light) gray. 
The value $R_{\tau/\mu}>1800$ means $\mathcal B (\tau\to\mu\gamma) > 10^{-9}$ for $\mathcal B (\mu\to e\gamma) = 5.7\times 10^{-13}$, the current experimental bound.  
When we fix the Majorana CP violating phases as $\alpha_\nu = 0,\,\pm\pi$ and $\pm 2\pi$, the regions are limited as exhibited in the figures. 
The horizontal width of the gray region in Fig.~\ref{FIG:contD}~(a) is proportional to $1/ \sqrt{m_{\nu_2} m_{\nu_3}}$ and that in Fig.~\ref{FIG:contD}~(b) to $1/ \sqrt{m_{\nu_1} m_{\nu_2}}$. 
This explains the smaller region of enhancement in the IH case (b). 
From this analysis, we conclude that there still remains possibilities to obtain $\mathcal B(\tau\to\mu\gamma) > 10^{-9}$ 
in the degenerate case with the CP violating parameters of $c$ (or $a$), $\delta_\nu$ and $\alpha_\nu$.

There are several experimental studies on $\delta_\nu$ by the T2K~\cite{DeltaCPT2K}, MINOS~\cite{DeltaCPMINOS} and Super-Kamiokande~\cite{DeltaCPSK} collaborations. 
For example, the T2K collaboration has reported that 
the Dirac CP violating phase of $0.19\pi < \delta_\nu < 0.8\pi\,\, (-0.04\pi < \delta_\nu <\pi)$ is excluded at $90\%$ CL for the normal (inverted) hierarchy 
combining their result with measurements of $\sin^2 \theta_{13}$ from reactor experiments.

\subsubsection{Non-degenerate case}
\begin{figure}
\includegraphics[viewport=0 0 730 400,width=36em]{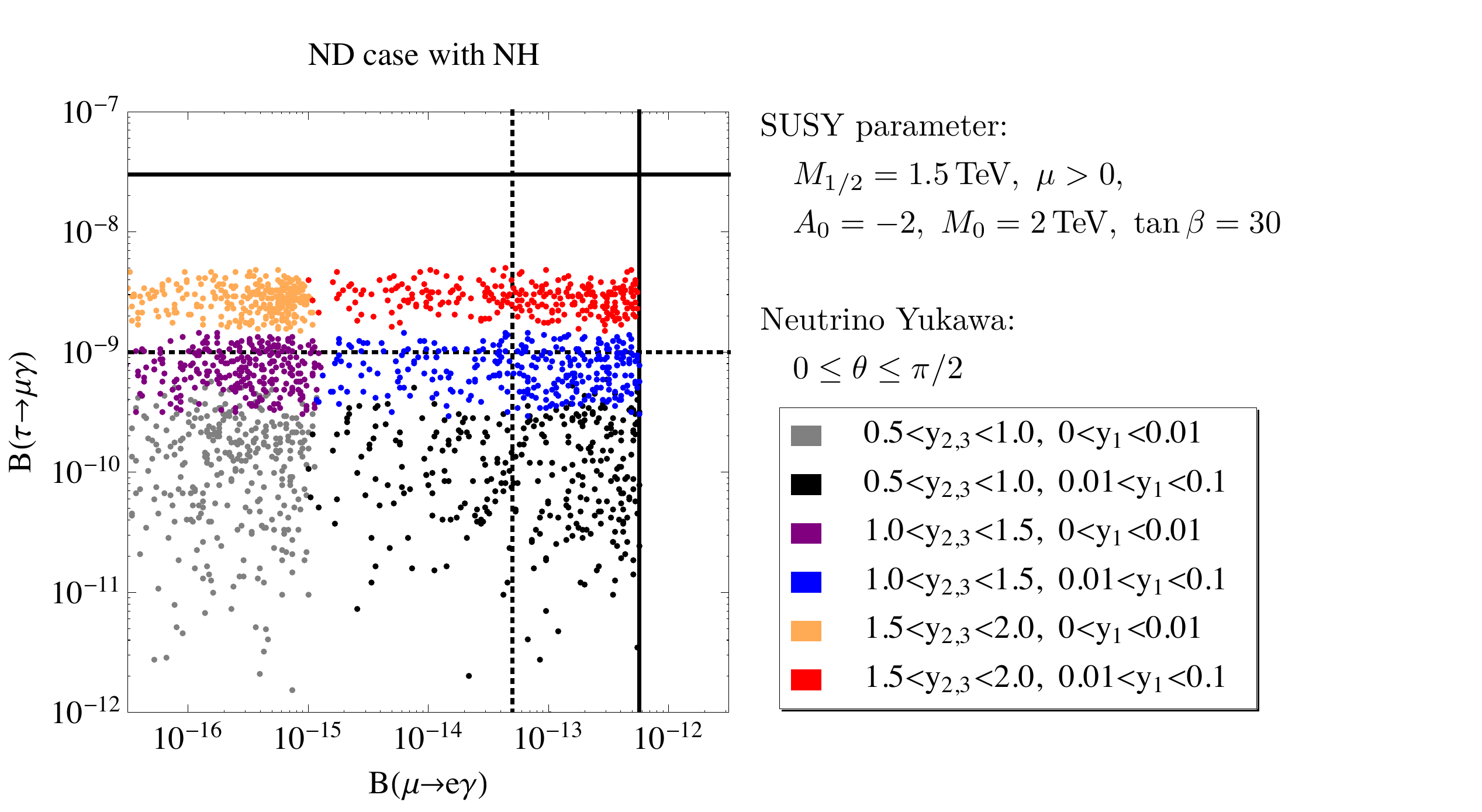} 
\caption{
Correlation between $\mathcal B(\mu \to e\gamma)$ and $\mathcal B(\tau\to\mu\gamma)$ in the non-degenerate case.   
The parameters $(y_1,y_2,y_3,\theta)$ in the $Y_N$ are randomly generated within the designated range 
and the SUSY parameters are fixed as $M_{1/2}=1.5\,\text{TeV},\,M_0=2.0\,\text{TeV},\,A_0=-2,\,\tan\beta=30$ and $\mu>0$. 
The solid lines show the present upper bounds of $\mathcal B(\mu \to e\gamma)$ and $\mathcal B(\tau\to\mu\gamma)$. 
The dotted lines indicate expected sensitivities at SuperKEKB/Belle~II and MEG~II. 
}
\label{FIG:scattND}
\end{figure}
We investigate $\mu\to e\gamma$ and $\tau\to\mu\gamma$ in the non-degenerate case. 
In order to simplify the following analysis, we assume the normal hierarchy for the neutrino masses and no CP violating phases in the lepton sector, namely $\delta_\nu = \alpha_\nu = \beta_\nu =0$. 
As explained in the previous section, we apply the parametrization in Eq.~(\ref{Eq:NDcaseSt}) to this case. 
In Fig.~\ref{FIG:scattND}, we show a scatter plot that represents the correlation between $\mathcal B(\tau\to\mu\gamma)$ and $\mathcal B(\mu\to e\gamma)$. 
In this plot, we explore the parameter space of $(y_1,y_2,y_3,\theta)$ in $Y_N$, fixing SUSY parameters as 
$M_{1/2}=1.5\,\text{TeV}$, $M_0=2.0\,\text{TeV}$, $A_0=-2$, $\tan\beta=30$ and $\mu>0$.
These parameters satisfy the constraints in Table~\ref{Tab:Bounds}. 
For the parameters $y_1$, $y_2$ and $y_3$, we divide their regions as indicated in the figure to show the dependence on them.  
The mixing angle between the second and third generation is taken in the region $0 \leq \theta \leq \pi/2$. 
It can be seen that $\mathcal B(\mu \to e\gamma)$ is sensitive to $y_1$ and $\mathcal B(\tau\to\mu\gamma)$ depends on $y_2$ and $y_3$. 
Large values of $y_2$ and $y_3$ are required for the enhancement of $\mathcal B(\tau\to\mu\gamma)$. 
The branching ratios for both tau and muon LFV decays can be large enough to be observed in near future.

\begin{figure}[th]
\includegraphics[viewport=0 0 730 400,width=36em]{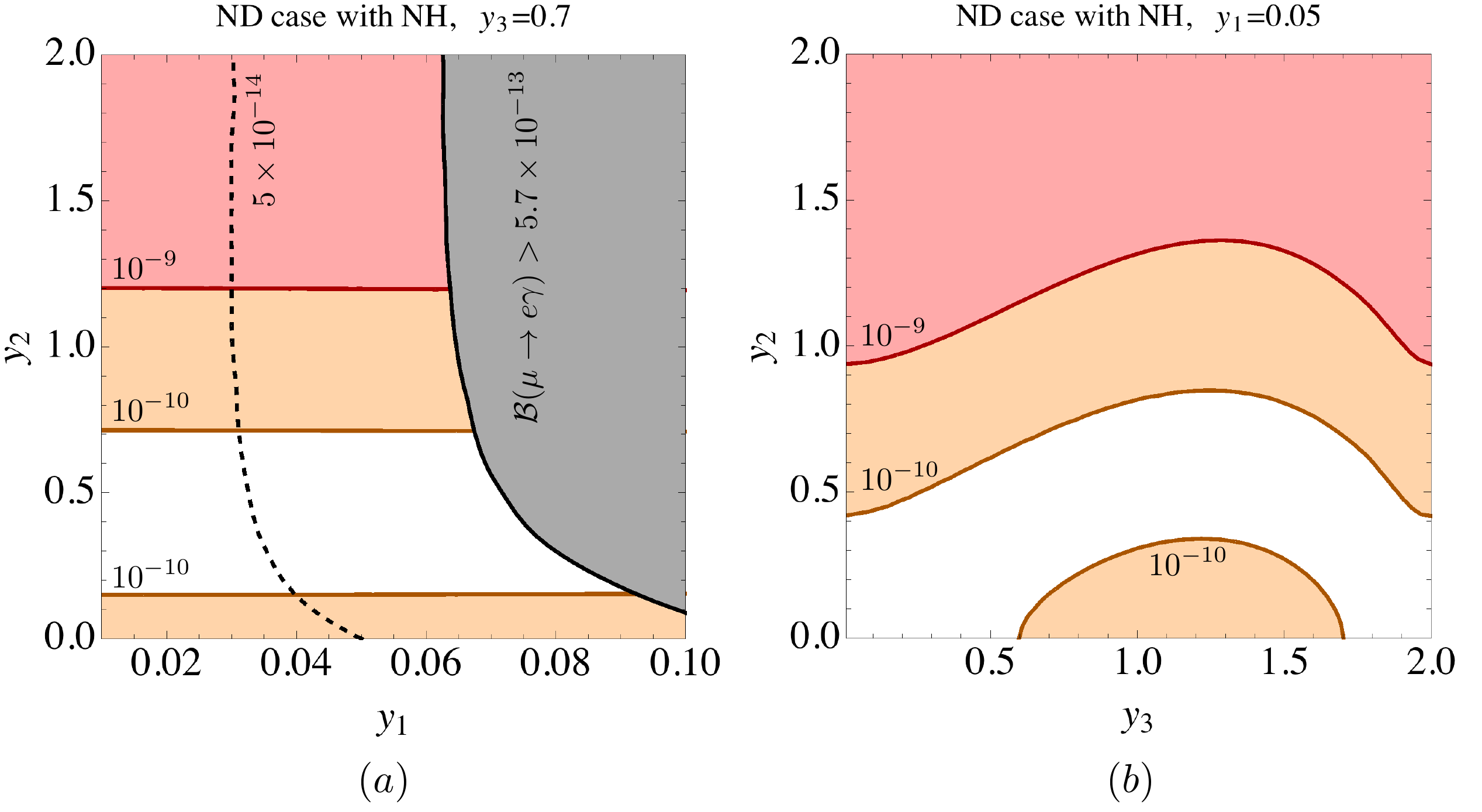} 
\caption{
Contour plots of $\mathcal B(\tau\to\mu\gamma)$ on planes of parameters in the neutrino Yukawa matrix in the non-degenerate case with the normal hierarchy and $\theta=\pi/4$.  
The red (orange) lines are $\mathcal B(\tau \to \mu\gamma) =10^{-9}$ ($10^{-10}$). 
The gray region is excluded by the present bound of $\mathcal B(\mu \to e\gamma)$. 
The dotted line shows the expected sensitivity of MEG~II, $\mathcal B(\mu \to e\gamma)=5\times 10^{-14}$. 
}
\label{FIG:contND}
\end{figure}
In order to see this situation more clearly,  
we show the contour plots of $\mathcal B(\tau\to\mu\gamma)$ 
on planes of $(y_1,y_2)$ in Fig.~\ref{FIG:contND}~(a) and $(y_3,y_2)$ in Fig.~\ref{FIG:contND}~(b) with $\theta=\pi/4$.  
We take $y_3=0.7$ in Fig.~\ref{FIG:contND}~(a) and $y_1=0.05$ in Fig.~\ref{FIG:contND}~(b).
In the red (orange) regions, $\mathcal B(\tau\to\mu\gamma)$ is larger than $10^{-9}$ ($10^{-10}$). 
We also show $\mathcal B(\mu \to e\gamma)$ in (a). 
The gray region is excluded by the current experimental upper bound of $\mathcal B(\mu \to e\gamma)$. 
In the whole region in Fig.~\ref{FIG:contND}~(b), $\mathcal B(\mu \to e\gamma) \simeq 3\times 10^{-13}$. 
As shown in Fig.~\ref{FIG:contND}, $\mathcal B(\mu \to e\gamma)$ is larger for larger $y_1$ and $\mathcal B(\tau\to\mu\gamma)$ mainly depends on $y_2$.

\begin{figure}[ht]
\includegraphics[viewport=0 0 730 400,width=36em]{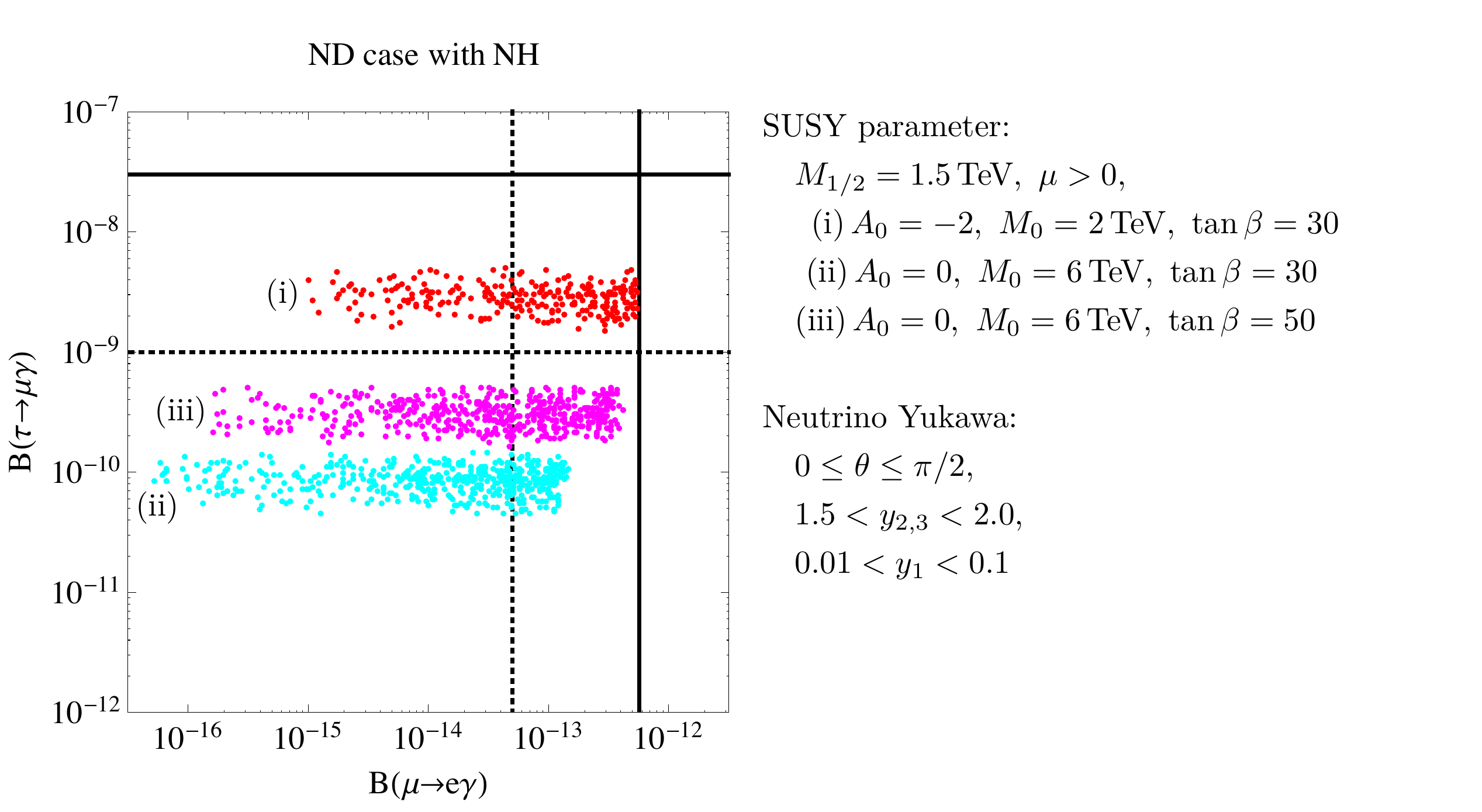} 
\caption{
Correlation between $\mathcal B(\mu \to e\gamma)$ and $\mathcal B(\tau\to\mu\gamma)$ in the non-degenerate case for three different choices of SUSY parameters.   
The parameters $(y_1,y_2,y_3,\theta)$ are randomly generated within the designated range. 
}
\label{FIG:scattNDzero}
\end{figure}
In Fig.~\ref{FIG:scattNDzero} we present a scatter plot of $\mathcal B(\mu \to e\gamma)$ and $\mathcal B(\tau\to\mu\gamma)$ for different choices of SUSY parameters:  
\begin{itemize}
\item[(i)] $A_0=-2,\,M_0=2\,\text{TeV}$ and $\tan\beta =30$, 
\item[(ii)] $A_0=0,\,M_0=6\,\text{TeV}$ and $\tan\beta =30$,  
\item[(iii)] $A_0=0,\,M_0=6\,\text{TeV}$ and $\tan\beta =50$,  
\end{itemize}
with $M_{1/2}=1.5\,\text{TeV}$ and $\mu >0$, where the parameters in the neutrino Yukawa matrix are randomly varied as indicated. 
These choices satisfy the experimental constraints and give $m_h \simeq 126\,\text{GeV}$. 
The choice (i) is the same as the plot in Fig.~\ref{FIG:scattND}. 
For the choice (ii), $A_0$ is set to zero and thus the large value of $M_0=6\,\text{TeV}$ is required to reproduce the observed Higgs boson mass. 
Accordingly both $\mathcal B(\mu \to e\gamma)$ and $\mathcal B(\tau\to\mu\gamma)$ are suppressed because of larger slepton masses. 
For the choice (iii), the large $\tan\beta$ increases $\mathcal B(\mu \to e\gamma)$ and $\mathcal B(\tau\to\mu\gamma)$, and 
the latter can be as large as $5 \times 10^{-10}$ within $y_2, y_3 <2.0$. 
We note that $y_2, y_3 \simeq 2.0$ gives the value of the heaviest right-handed neutrino mass close to the GUT scale. 
In all these cases, $\mathcal B(\mu \to e\gamma)$ can be larger than $5 \times 10^{-14}$ if $y_1$ is close to $0.1$.

\section{Summary and conclusion}
\label{Co}
We have studied the lepton flavor violation in the supersymmetric seesaw model of type~I with the ansatz from the minimal supergravity. 
We have evaluated the latest constraints on the SUSY parameters, taking into account recent experimental improvements 
for the Higgs boson mass and direct searches of the SUSY particles at the LHC, the rare decay of $B_s \to \mu^+\mu^-$ at the dedicated $B$ experiments, 
the neutrino mixing angle of $\theta_{13}$ at the neutrino experiments and the charged lepton flavor violating decay at the MEG experiment.   
The Higgs boson mass strongly constrains the SUSY parameters and 
we have shown that the allowed region of the universal scalar mass $M_0$ and gaugino mass $M_{1/2}$ depends on the universal trilinear coupling $A_0$ as shown in Figs.~\ref{FIG:constD} and \ref{FIG:constND}. 
We have also found that the constraints from the $B$ physics are important because $\mathcal B(\bar B\to X_s\gamma)$ gives the strong constraint in a certain parameter region.

Using the obtained allowed region of the SUSY parameters, 
we have investigated the effect of the parameters in the neutrino Yukawa matrix $Y_N$ and Majorana mass matrix $M_N$ on the LFV decays $\tau\to\mu\gamma$ and $\mu\to e\gamma$. 
In this study, we considered the degenerate and non-degenerate cases for $Y_N$ and $M_N$. 
In the degenerate case, $M_N$ is assumed to be proportional to the unit matrix.  
We have found that $\mathcal B(\tau\to\mu\gamma)$ is less than $2 \times 10^{-12}$ for $O_N={\bf 1}$ 
with the present bound of $\mathcal B(\mu \to e\gamma)$ and the current experimental value of $\sin^2\theta_{13}$.  
However, $\mathcal B(\tau\to\mu\gamma)$ can be larger than $10^{-9}$ 
when the CP violating parameters in $O_N$ are taken into account together with the CP violating phases in $U_\nu$. 
In the non-degenerate case, we assume that $Y_N$ has a mixing only between the second and third generations. 
In this case, $\mathcal B(\mu \to e\gamma)$ and $\mathcal B(\tau\to\mu\gamma)$ depend on the different parameters in $Y_N$.   
We have found that $\mathcal B(\tau\to\mu\gamma)$ can be larger than $10^{-9}$ for $A_0=-2$. 
For a smaller value of $|A_0|$, the branching ratio is smaller because of the required large masses of sleptons.

The future experiment at SuperKEKB/Belle~II is able to search for $\tau\to\mu\gamma$ down to $\mathcal B(\tau\to\mu\gamma) \sim 10^{-9}$~\cite{Aushev:2010bq}. 
In our analysis $\mathcal B(\tau\to\mu\gamma)>10^{-9}$ can be obtained for both degenerate and non-degenerate cases. 
For the search for muon LFV processes, several new and upgraded experiments are now under construction. 
The MEG~II experiment~\cite{Baldini:2013ke} can reach a sensitivity down to $\mathcal B(\mu\to e\gamma)\simeq 5\times 10^{-14}$. 
The phase~II COMET experiment~\cite{Cui:2009zz} and the Mu2e experiment~\cite{Abrams:2012er} can reach to $\mathcal B(\mu^-N \to e^-N)\sim 10^{-17}$-$10^{-18}$, 
which corresponds to $\mathcal B(\mu\to e\gamma) \sim 10^{-14}$-$10^{-15}$ in the SUSY seesaw model where the photon dipole operator gives dominant contributions.  
Thus, the future experiments for the $\mu$-$e$ conversion search will be also useful to investigate the SUSY seesaw model.

In conclusion, the SUSY seesaw model in the context of the minimal supergravity is a viable new physics candidate that is consistent with the observed Higgs boson mass and other new physics searches. 
The exploration of LFV processes at the intensity frontier may provide us with the signal of the SUSY seesaw model in conjunction with the new physics search at the energy frontier.


\begin{acknowledgments}
We are grateful to Werner Porod for a helpful comment on the renormalization group running in the public code of {\it SPheno}. 
This work is supported in part by JSPS KAKENHI, Grant Numbers 20244037 and 22244031 for YO, 23104011 and 24340046 for TS, 25400257 for MT 
and by IBS-R018-D1 for RW.  
\end{acknowledgments}


\end{document}